\documentclass{Interspeech}
\usepackage{graphicx}
\usepackage{subcaption}
\usepackage{multirow}
\usepackage{lipsum}
\usepackage[table,xcdraw]{xcolor}
\usepackage{amssymb}
\usepackage{colortbl}
\usepackage{xcolor}
\usepackage{bbm}
\usepackage{subfiles}
\usepackage{MnSymbol}

\definecolor{armygreen}{rgb}{0.29, 0.33, 0.13}
\definecolor{forestgreen}{rgb}{0.13, 0.54, 0.13}
\newcommand{\parlabel}[1]{\vspace{0em}\noindent\textbf{#1}.}

\def\ourapproach{\textsc{QUADS}\xspace}

\newcommand\blfootnote[1]{%
  \begingroup
  \renewcommand\thefootnote{}\footnote{#1}%
  \addtocounter{footnote}{-1}%
  \endgroup
}

\interspeechcameraready

\title{\ourapproach: \textsc{Qua}ntized \textsc{D}istillation Framework for Efficient \textsc{S}peech Language Understanding}

\author[affiliation={*}]{Subrata}{Biswas}
\author[affiliation={*}]{Mohammad Nur Hossain}{Khan}
\author[]{Bashima}{Islam}
\affiliation{}{Worcester Polytechnic Institute}{USA}
\email{\texttt{\{sbiswas, mkhan, bislam\}@wpi.com}}

\usepackage{comment}

\begin{document}

\maketitle

\begin{abstract}

Spoken Language Understanding (SLU) systems must balance performance and efficiency, particularly in resource-constrained environments. Existing methods apply distillation and quantization separately, leading to suboptimal compression as distillation ignores quantization constraints. We propose \ourapproach, a unified framework that optimizes both through multi-stage training with a pre-tuned model, enhancing adaptability to low-bit regimes while maintaining accuracy. \ourapproach achieves 71.13\% accuracy on SLURP and 99.20\% on FSC, with only minor degradations of up to 5.56\% compared to state-of-the-art models. Additionally, it reduces computational complexity by 60–73$\times$ (GMACs) and model size by 83–700$\times$, demonstrating strong robustness under extreme quantization. These results establish \ourapproach as a highly efficient solution for real-world, resource-constrained SLU applications.


\end{abstract}

\blfootnote{\textsuperscript{$*$}Equal contribution. \\This work is supported by the NSF grant CNS-2347692.}
\section{Introduction}

Spoken Language Understanding (SLU), a critical component of Natural Language Understanding, aims to extract semantic information from user utterances \cite{cheng2023ml} and intent detection is one of the key subtasks of SLU, involving classifying the overall purpose of an utterance. As Augmented Reality (AR), Virtual Reality (VR), and voice-assisted technologies continue to proliferate, SLU has become a cornerstone of conversational AI, enabling systems to interpret and derive meaning from user input \cite{zhu2024zero, rajaa23_interspeech, cheng2024towards, zhuang2024towards}. With the growing ubiquity of these technologies in everyday life—from virtual assistants like Alexa and Siri to immersive AR applications—the demand for SLU systems that are not only accurate but also responsive, secure, and efficient has never been greater.

Conventional SLU systems typically follow a two-stage process: first, an Automated Speech Recognition (ASR) module converts spoken audio into text; then, an SLU module analyzes the transcribed text to detect the user's intent \cite{kim2023efficient, cheng2024moe, cheng2023c}.  Recent works integrate large language models (LLMs) with ASR for intent classification, yielding promising results \cite{zhu2024zero, hoscilowicz2024large, cho2024zero, fan2024lanid}. However, these architectures are vulnerable to error propagation, where transcription inaccuracies from the ASR module adversely impact intent classification performance \cite{tran2020neural, rajaa23_interspeech, wei2022neural, wallbridge2023dialogue, cheng2024moe}. To address this limitation, researchers have explored end-to-end models that directly classify intent from speech, bypassing the intermediate transcription step \cite{wei2022neural, rajaa23_interspeech, zhuang2024towards, wang2023whislu, kim2024joint, chen2021top}. These models achieve high accuracy and are particularly suitable for AR, VR, and voice-assisted devices.

\begin{figure*}[!ht]
\begin{minipage}{0.64\textwidth}
    \centering  
    \includegraphics[width=\textwidth]{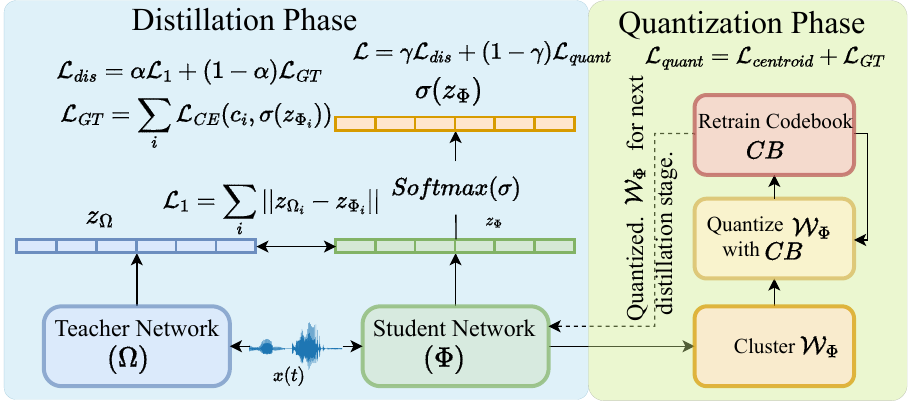}
    \vspace{-2em}
    \caption{\textbf{Schematic overview of \ourapproach.}  A two-phase framework for efficient model training. In the distillation phase, the student model $\Phi$ learns from the teacher model $\Omega$ via a combined loss strategy. The quantization phase compresses the student model's weights $\mathcal{W}_{\Phi}$ using the codebook,  where weights are grouped into clusters and refined using objectives that balance centroid alignment and cross-network consistency.}
    \label{main_diagram}
\end{minipage}
\hspace{0.5em}
\begin{minipage}{0.35\textwidth}
    \centering  
    \includegraphics[width=\textwidth]{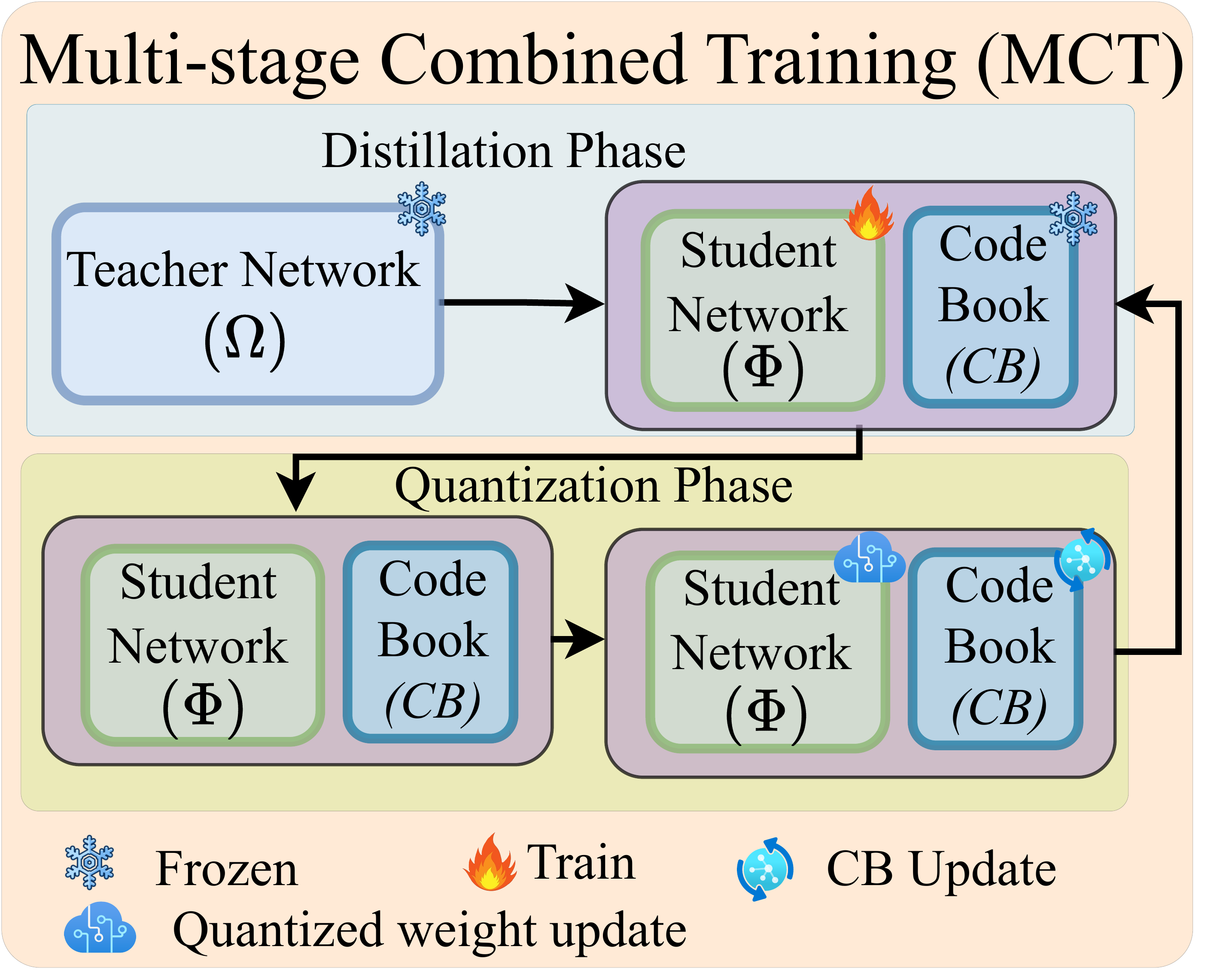}
    \vspace{-2em}
    \caption{\textbf{Training stages of \ourapproach.} The distillation phase transfers knowledge to the student, and in the quantization phase, the distilled student undergoes quantized weight updates.}
    \label{training_stages}
\end{minipage}
\end{figure*}

However, the substantial size of these end-to-end models—typically ranging from 75.53 to 2422 MB—poses significant challenges for on-device deployment. As a result, these models are often processed in the cloud, introducing practical issues such as increased latency, elevated energy consumption, and privacy risks. For instance, in AR and VR environments, where immediate feedback is crucial for user immersion, latency can disrupt the experience \cite{kohrs2016delays, porcheron2018voice}. Similarly, in voice-assisted devices handling sensitive information, cloud processing raises privacy concerns \cite{sun2020alexa}. Frequent data transmission between devices and the cloud also leads to higher energy consumption, impacting battery life in portable devices \cite{alexandridis2022tinys2i, benazir2024speech}.

These limitations highlight the pressing need for lightweight, efficient SLU models capable of on-device processing. While model compression techniques like knowledge distillation ~\cite{gou2021knowledge}, quantization ~\cite{wu2016quantized}, and pruning \cite{wang2020pruning} have been widely adopted to reduce model size, they often fall short in preserving performance due to their sequential application. Traditional methods typically pre-train models using distillation and then apply quantization, leading to compounded errors and suboptimal compression \cite{li2022dq, zhang2020ternarybert}. Disjoint distillation and quantization stages introduce error propagation, where information loss during distillation compounds during quantization, degrading overall performance. Furthermore, this approach struggles with low-bit quantization, as distilled models are not inherently adapted to extreme precision constraints, resulting in significant quantization errors.

To overcome these challenges, this paper introduces \ourapproach, a unified \textsc{Qua}ntized \textsc{D}istillation framework explicitly designed for Spoken Language Understanding (SLU) tasks. By seamlessly integrating distillation and quantization into a cohesive process, \ourapproach addresses the limitations of traditional methods and enables efficient, high-performance SLU deployment on resource-constrained devices.

Our contributions tackle three critical challenges in developing efficient SLU models for on-device deployment:

\noindent \textbf{1. Unified Distillation and Quantization for Efficient Compression:}
Balancing model compression with performance retention is a significant challenge, as reducing model size often leads to degraded accuracy. \ourapproach mitigates this issue by integrating distillation and quantization into a unified process, preventing the error propagation that occurs when these steps are treated separately. This cohesive framework preserves model performance even under extreme compression, enabling robust SLU in compact models.

\noindent \textbf{2. Adaptability to Low-Bit Quantization:}
Adapting models to low-bit quantization without compromising intent detection accuracy is another key challenge. \ourapproach employs a multi-stage combined training strategy that concurrently optimizes the model for both distillation and quantization. This joint optimization enhances the model's adaptability to extreme precision reductions, ensuring high performance even in low-bit regimes.

\noindent \textbf{3. Robust Generalization in Diverse Acoustic Environments:}
Compressed models often struggle to generalize well across diverse and noisy acoustic environments due to reduced capacity. To address this, \ourapproach leverages pretrained acoustic-linguistic representations, enhancing robustness and maintaining high accuracy despite aggressive compression. This ensures consistent performance across varied real-world SLU scenarios.
\section{\ourapproach: Quantized Distillation}
\label{quantized_distillation}



 To achieve maximum accuracy with minimum computational complexity for a given speech signal $x(t)$ for intent classification, we adopt an Expectation Maximization (EM)\cite{em} approach by integrating model distillation and quantization into a cohesive framework.(Fig.\ref{main_diagram}). Our proposed quantized distillation (QD) framework leverages an iterative multi-stage combined training procedure (MCT) (Fig.~\ref{training_stages}) to achieve this balance. The following section details critical phases of this process along with the MCT pipeline.

\begin{table*}[]
\caption{\textbf{Comparison of \ourapproach and Prior Methods on the SLURP and FSC Datasets.} We report accuracy and F1-score for model performance, alongside GMACs and model size, to evaluate efficiency.}
\label{tab:main_result}
\resizebox{0.99\textwidth}{!}{
\begin{tabular}{l|r|rrrrr|rrrrr}
\hline
\multicolumn{1}{c|}{}                           & \multicolumn{1}{l|}{}                             & \multicolumn{5}{c|}{SLURP}                                                                                                                                                                       & \multicolumn{5}{c}{FSC}                                                                                                                                         \\ \cline{3-12} 
\multicolumn{1}{c|}{\multirow{-2}{*}{Baseline}} & \multicolumn{1}{l|}{\multirow{-2}{*}{\begin{tabular}[c]{@{}c@{}}Bit\\Length\end{tabular}}} & \multicolumn{1}{c}{\begin{tabular}[c]{@{}c@{}}\#Param\\(M) $\downarrow$\end{tabular}} & \multicolumn{1}{c}{\begin{tabular}[c]{@{}c@{}}Model\\Size (MB) $\downarrow$\end{tabular}} & \multicolumn{1}{c}{GMACs $\downarrow$} & \multicolumn{1}{c}{Accuracy $\uparrow$}                 & \multicolumn{1}{c|}{F1-Score $\uparrow$}                 & \multicolumn{1}{c}{\begin{tabular}[c]{@{}c@{}}\#Param\\(M) $\downarrow$\end{tabular}} & \multicolumn{1}{c}{\begin{tabular}[c]{@{}c@{}}Model\\Size (MB) $\downarrow$\end{tabular}} & \multicolumn{1}{c}{GMACs $\downarrow$} & \multicolumn{1}{c}{Accuracy $\uparrow$} & \multicolumn{1}{c}{F1-Score $\uparrow$} \\ \hline
CTL\textsubscript{pt}                                      & 32                                                & 127                             & 484.47                              & 143.2                     & 90.14                        & 82.27                         & \multicolumn{1}{c}{-}           & \multicolumn{1}{c}{-}               & \multicolumn{1}{c}{-}     & \multicolumn{1}{c}{-}        & \multicolumn{1}{c}{-}        \\
CTL                                             & 32                                                & 127                             & 484.47                              & 143.2                     & 72.56                        & 43.34                         & \multicolumn{1}{c}{-}           & \multicolumn{1}{c}{-}               & \multicolumn{1}{c}{-}     & \multicolumn{1}{c}{-}        & \multicolumn{1}{c}{-}        \\
Whisper (large)                                 & 32                                                & 634.94                          & 2422.10                             & 1136.58                   & 75.32                         & 71.11                          & 634.90                          & 2421.97                             & 1136.58                   & 99.49                        & 99.44                        \\
Whisper (small)                                 & 32                                                & 87.05                           & 332.1                               & 172.14                    &  72.16                        & 69.73                         & 87.03                           & 331.97                              & 172.13                    & 99.39                        & 99.31                        \\
Whisper (base)                                  & 32                                                & 19.85                           & 75.73                               & 43.71                     & 71.7                         & 65.48                         & 19.84                           & 75.68                               & 43.70                      & 99.44                        & 99.4                         \\
Prosody                                         & 32                                                & 21.04                           & 80.27                               & 43.82                     & 68.23                        & 62.55                         & 21.04                           & 80.27                               & 44.12                     & 97.80                         & 98.10                         \\
Prosody + Distillation                          & 32                                                & 21.47                           & 81.92                               & 44.09                     & 76.26                        & 71.92                         & 21.47                           & 81.92                               & 44.31                     & 99.10                         & 98.30                         \\ \hline
\rowcolor[HTML]{E7FAE6} 
\cellcolor[HTML]{E7FAE6}                        & 32                                                & 7.25                            & 27.66                               & 15.6                      & 71.13                        & 65.07                         & 7.64                            & 29.16                               & 18.48                     & 99.20                         & 99.10                           \\
\rowcolor[HTML]{E7FAE6} 
\cellcolor[HTML]{E7FAE6}                        & 16                                                & 7.25                            & 13.83                               & 15.6                      & 70.48                        & 65.21                         & 7.64                            & 14.58                               & 18.48                     & 98.78                        & 98.21                        \\
\rowcolor[HTML]{E7FAE6} 
\cellcolor[HTML]{E7FAE6}                        & 8                                                 & 7.25                            & 6.91                                & 15.6                      & 69.73                        & 64.87                         & 7.64                            & 7.29                                & 18.48                     & 98.20                         & 97.87                        \\
\rowcolor[HTML]{E7FAE6} 
\multirow{-4}{*}{\cellcolor[HTML]{E7FAE6}\ourapproach}  & 4                                                 & 7.25                            & 3.46                                & 15.6                      & 68.98                        & 64.39                         & 7.64                            & 3.65                                & 18.48                     & 97.39                        & 96.12                        \\ \hline
\end{tabular}
}
\vspace{-1em}
\end{table*}

\subsection{Distillation Phase}
The QD process begins with extracting the mel-spectrogram $X(t, f) = f(x(t))$ from the speech signal $x(t)$, where $f(\cdot)$ represents the mel-spectrogram extraction function. This spectrogram serves as input for both the highly capable, computationally expensive teacher model $\Omega$ and the lightweight student network $\Phi$.

The feature representations from the teacher and student networks are denoted as $z_{\Omega} = f_{\Omega}(X(t, f)) \in \mathbb{R}^n$ and $z_{\Phi} = f_{\Phi}(X(t, f)) \in \mathbb{R}^n$, where $f_{\Omega}(\cdot)$ and $f_{\Phi}(\cdot)$ are the respective feature encoders, and $n$ is the latent space size for both networks. To align the student network's feature space $z_{\Phi}$ with the teacher's $z_{\Omega}$, we compute the $l_1$ loss:

\begin{equation} \label{l1_loss} \mathcal{L}_1 = \sum{(z_{\Omega_i}, z_{\Phi_i})} || z_{\Omega_i} - z_{\Phi_i} ||_1 \end{equation}

A classification head is appended to the student network’s encoder $f_{\Phi}(\cdot)$ to perform intent classification. The cross-entropy loss between the student network's predictions and the ground truth labels is defined as:

\begin{equation} \label{gt_loss} \mathcal{L}_{GT} = \sum_i \mathcal{L}_{CE}(c_i, \sigma(z_{\Phi_i})) \end{equation}

Here, $\sigma(\cdot)$ represents the \textit{softmax} function, and $c_i$ denotes the ground truth labels.
The combined distillation loss $\mathcal{L}_{dis}$ balances feature alignment and classification accuracy:

\begin{equation} \label{dis_loss} \mathcal{L}_{dis} = \alpha\mathcal{L}_{1} + (1-\alpha)\mathcal{L}_{GT} \end{equation}

\subsection{Quantization Phase.}
Following distillation, the student model's weights $\mathcal{W}_{\Phi}$ are quantized using a $k$-means clustering-based method to further reduce computational complexity. For a given bit length $b$, $k = 2^b$ centroids are initialized randomly, and the weights $\mathcal{W}_{\Phi}$ are partitioned into $k$ clusters by minimizing the within-cluster sum of squares:
$\mathop{\arg \min}\limits_{C} \sum_{i=1}^k \sum_{w \in c_i} |w - c_i|^2$

During back-propagation, the gradient of each weight is calculated to update the centroids~\cite{han2015deep}. The centroid update loss $\mathcal{L}_{centroid}$ is defined using the indicator function $\mathbbm{1}(\cdot)$:

\begin{equation} \frac{\partial \mathcal{L}_{centroid}}{\partial C_k} = \sum_{i, j} \frac{\partial \mathcal{L}_{centroid}}{\partial \mathcal{W}_{\Phi_{i,j}}} \mathbbm{1}(I_{i,j} = k) \end{equation}

The total quantization loss $\mathcal{L}_{quant}$ is defined as:

\begin{equation} \label{quant_loss} \mathcal{L}_{quant} = \mathcal{L}_{centroid} + \mathcal{L}_{GT} \end{equation}

To effectively integrate both distillation and quantization, we employ a multi-stage training strategy described below.

\subsection{Multi-stage Combined Training}
\label{multistage_training}

To transfer knowledge effectively while maintaining a compact model, we employ a \underline{M}ulti-Stage \underline{C}ombined \underline{T}raining (MCT) strategy, grounded in the principles of Expectation Maximization (EM). MCT alternates between distillation and quantization phases, treating distillation as the expectation step, where the student model learns from the teacher, and quantization as the maximization step, where model parameters are optimized for efficiency. This iterative process progressively refines the student model, as illustrated in Figure~\ref{training_stages}.
The total loss $\mathcal{L}$ is computed by combining the distillation and quantization losses:

\begin{equation} \mathcal{L} = \gamma \mathcal{L}_{dis} + (1 - \gamma) \mathcal{L}_{quant} \end{equation}

Here, $\gamma \in \{0, 1\}$ controls the training phase, with $\gamma = 1$ during distillation and $\gamma = 0$ during quantization.

After multiple cycles of distillation and quantization, a final quantization phase is applied. This ensures the model is optimally compressed while maintaining high performance. Unlike intermediate quantization steps, the final phase focuses exclusively on minimizing the model footprint for deployment in resource-constrained environments, solidifying the student model’s ability to operate without significant loss in accuracy.

\section{Experiments}

\subsection{Dataset}
Following the evaluation of the latest works on intent classification~\cite{rajaa23_interspeech, zhu2024zero, chen2021top}, we conduct experiments on two prominent SLU datasets, SLURP and FSC, to ensure comprehensive evaluation across diverse domains and command-specific tasks.

\noindent{\textbf{{SLURP. \cite{bastianelli2020slurp}}}} The dataset comprises $72$K $16$kHz spoken-language-understanding recordings across 18 distinct domains, split into 49.9K (39.7 h) train, 8.5K (6.8 h) validation, and 12.9K (10.1 h) test utterances.


\noindent{\textbf{FSC.}}\cite{lugosch2019speech}The dataset consists of 30,000 16kHz single-channel audio recordings of English commands from 97 distinct users designed for smart home and virtual assistant applications. 

\subsection{State-of-the-Art (SOTA) Baseline Models}

We compare \ourapproach against several SOTA models equipped with either SLU or distillation frameworks \cite{gulati2020conformer, radford2023robust, rajaa23_interspeech} while varying model sizes to understand its scalability and efficiency.

\noindent\textbf{Conformer-Transformer-Large (CTL)~\cite{gulati2020conformer}} employs a transformer architecture, leveraging convolutional modules to capture local temporal features and transformer modules to model global dependencies in the audio signal.

\noindent\textbf{Whisper~\cite{radford2023robust}} uses convolutional blocks to extract features from the log–mel spectrograms and then passes the feature through a transformer architecture to generate text in an autoregressive manner. We evaluate against the \textit{small}, \textit{base}, and \textit{large} variants.

\noindent\textbf{Prosody~\cite{rajaa23_interspeech}} leverages prosodic features to generate an attention map for audio over time. We compare against both \textit{prosody-only} and a distillation enhanced \textit{(prosody + distillation)}.

\subsection{Evaluation Metrics}
To evaluate model performance, we report both \textit{accuracy} and \textit{F1-score} on the test sets. 
For model efficiency and computational complexity, we report the number of parameters, model size (in megabytes, MB), and the number of multiplication and accumulation operations (GMACs) during inference.

\subsection{Implementation Details}

We use Whisper \textit{large} as our teacher model $\Omega$. Since the Whisper model accepts mel spectrograms as inputs, we compute an 80-channel log mel spectrogram for all speech samples using 25-millisecond windows with a 10-millisecond stride. Our student model $\Phi$ follows a structure similar to that of the teacher. Further implementation details can be found in the open-source codebase \footnote{\url{https://github.com/BASHLab/QUADS}}. 
Our student model's encoder is initialized with Whisper pre-trained weights. We add a classifier head after the encoder of $\Phi$ for intent classification.


To train the student model, we use learning rates of $1 \times 10^{-6}$ for the encoder and $1 \times 10^{-3}$ for the classifier. 
Our iterative multi-stage training alternates between distillation and quantization for five iterations, with each phase running for epochs.
\begin{figure*}[!ht]
\begin{minipage}{0.59\textwidth}
        \centering
    \captionsetup{type=figure}
    \captionsetup{position=bottom}
    \resizebox{0.99\textwidth}{!}{
    \begin{tabular}{c}
        \includegraphics[width=0.99\textwidth]{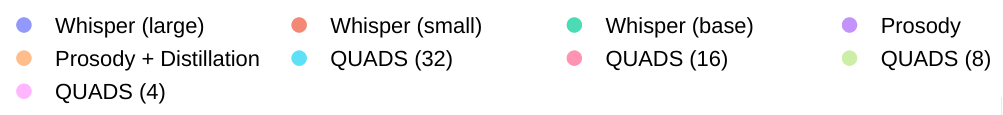} \\
           \subfloat[SLURP
      \label{bubble_slurp}]{\includegraphics[width=.49\linewidth]{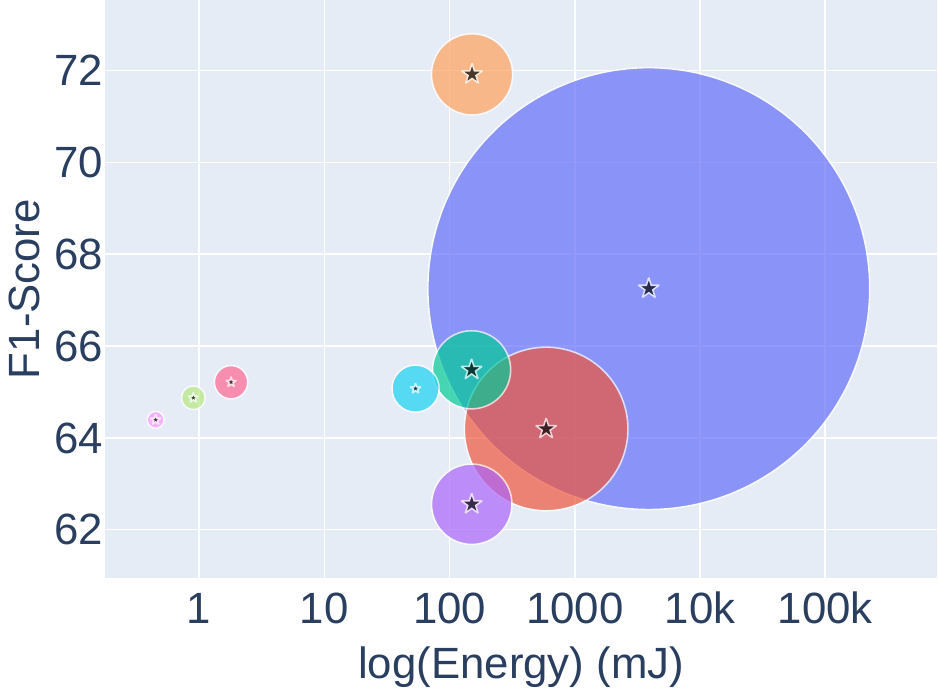}}
    \hspace{.1in}
    \subfloat[FSC
    \label{bubble_fsc}]{\includegraphics[width=.49\linewidth]{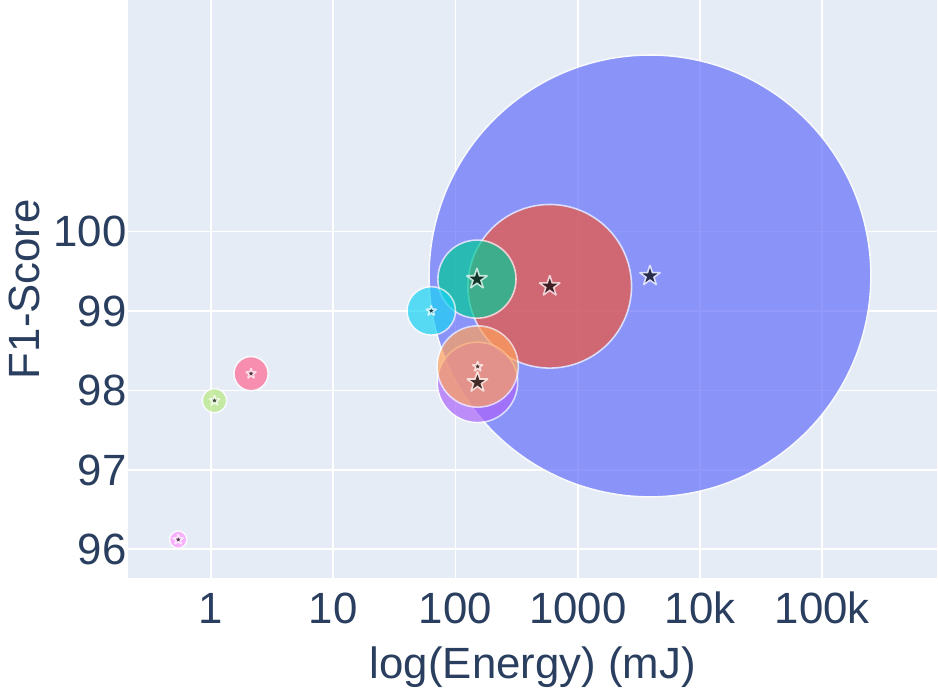}}
     \\
    \end{tabular}
    }
    \vspace{-1em}
    \caption{ 
        \ourapproach effectively shrinks model size to 3.46–29.16 MB, significantly reducing power consumption—by up to 700$\times$ for \textbf{SLURP} and 663$\times$ for \textbf{FSC}—while preserving competitive F1-scores, with a maximum drop of 5.56\%. In the visualization, bubble diameter represents model size (MB), and the center of each bubble ($\filledstar$) marks the F1-score at a given power consumption.}
    
    
\end{minipage}
\hspace{0.5em}
\begin{minipage}{0.4\textwidth}
\captionsetup{type=table}
\caption{\textbf{Ablation on model initialization and different training strategy.} We study the effect of model initialization and training methods on \ourapproach.}
\label{tab:ablation}
\resizebox{\linewidth}{!}{
\begin{tabular}{l|r|l|rr|rr}
\toprule
\multicolumn{1}{c|}{}                                 & \multicolumn{1}{c|}{}                             & \multicolumn{1}{c|}{}                                    & \multicolumn{2}{c|}{SLURP}                                     & \multicolumn{2}{c}{FSC}                                        \\ \cmidrule{4-7} 
\multicolumn{1}{c|}{\multirow{-2}{*}{\begin{tabular}[c]{@{}c@{}}Initia-\\lization\end{tabular}}} & \multicolumn{1}{c|}{\multirow{-2}{*}{\begin{tabular}[c]{@{}c@{}}Bit\\Length\end{tabular}}} & \multicolumn{1}{c|}{\multirow{-2}{*}{\begin{tabular}[c]{@{}c@{}}Training\\Strategy\end{tabular}}} & \multicolumn{1}{c}{\begin{tabular}[c]{@{}c@{}}Model\\Size (MB) $\downarrow$\end{tabular}} & \multicolumn{1}{c|}{\begin{tabular}[c]{@{}c@{}}F1-\\Score $\uparrow$\end{tabular}} & \multicolumn{1}{c}{\begin{tabular}[c]{@{}c@{}}Model\\Size (MB) $\downarrow$\end{tabular}} & \multicolumn{1}{c}{\begin{tabular}[c]{@{}c@{}}F1-\\Score $\uparrow$\end{tabular}}  \\ \midrule
                                                      &                                                   & Distillation                                        & 78.31                          & 26.59                         & 137.39                         & 88.93                         \\ \cmidrule{3-7}
                                                    \multirow{-1}{*}{Random}   &                                                  \multirow{-1}{*}{16} & \begin{tabular}[c]{@{}c@{}}Quantization\\after\\Distillation\end{tabular}                          & 48.13                          & 12.91                         & 72.31                          & 85.63                         \\ \cmidrule{3-7}
                            &                               & \cellcolor[HTML]{E7FAE6}MCT                              & \cellcolor[HTML]{E7FAE6}13.83  & \cellcolor[HTML]{E7FAE6}39.78 & \cellcolor[HTML]{E7FAE6}37.83  & \cellcolor[HTML]{E7FAE6}90.19 \\ \cmidrule{2-7}
                            & 4                                                 & \cellcolor[HTML]{E7FAE6}MCT                              & \cellcolor[HTML]{E7FAE6}3.46   & \cellcolor[HTML]{E7FAE6}29.61 & \cellcolor[HTML]{E7FAE6}9.4597 & \cellcolor[HTML]{E7FAE6}89.71 \\ \midrule
                                                      &                                                   &Distillation                                        & 78.31                          & 60.93                         & 137.39                         & 98.71                         \\ \cmidrule{3-7}
                                                      &                                                   & \begin{tabular}[c]{@{}c@{}}Quantization\\after\\Distillation\end{tabular}                          & 48.13                          & 53.79                         & 72.31                          & 96.23                         \\ \cmidrule{3-7}
                                                      & \multirow{-5}{*}{16}                              & \cellcolor[HTML]{E7FAE6}MCT                              & \cellcolor[HTML]{E7FAE6}13.83  & \cellcolor[HTML]{E7FAE6}65.21 & \cellcolor[HTML]{E7FAE6}37.83  & \cellcolor[HTML]{E7FAE6}98.21 \\ \cmidrule{2-7}
\multirow{-6}{*}{\begin{tabular}[c]{@{}c@{}}Pre-\\trained\end{tabular}}                         & 4                                                 & \cellcolor[HTML]{E7FAE6}MCT                              & \cellcolor[HTML]{E7FAE6}3.46   & \cellcolor[HTML]{E7FAE6}64.39 & \cellcolor[HTML]{E7FAE6}9.4597 & \cellcolor[HTML]{E7FAE6}96.12 \\ \bottomrule
\end{tabular}
}
\end{minipage}
\vspace{-2em}
\end{figure*}

\section{Results}


\subsection{Comparison with Baseline Algorithms}
Table~\ref{tab:main_result} presents a comprehensive comparison between \ourapproach and SOTA models on the SLURP and FSC datasets. \ourapproach consistently demonstrates superior efficiency and scalability, making it an ideal candidate for real-world, on-device applications without compromising performance.

\parlabel{SLURP Dataset}
On SLURP, \ourapproach achieves an \textit{F1-score} of $64.39$–$65.21$\% and accuracy of $68.98$–$71.13$\%, while drastically reducing computational overhead. With a minimal model footprint ranging from $3.46$~MB to $27.66$~MB and requiring only $15.60$ GMACs, \ourapproach achieves results that are highly competitive with larger, more resource-intensive models.

In contrast, SOTA baselines show marginally higher \textit{F1-scores} of $65.48$–$71.92$\% (an average of just $3.9\%$ improvement), but at a significant cost: they demand up to $3\times$ more computational resources (GMACs of $43.72$–$44.09$) and models that are $2.9$–$23\times$ larger (75.73 MB to 81.92 MB). This highlights \ourapproach’s unparalleled efficiency, delivering nearly equivalent performance with a fraction of the resource demands.
Notably, our $4$-bit quantized model occupies only $3.46$ MB and contains just $7.25$ million parameters while maintaining robust performance, with at most $7.53\%$ drop compared to the Whisper-distilled prosody model. This level of compression, paired with minimal performance degradation, underscores \ourapproach’s potential for deployment in resource-constrained environments.

Moreover, while all versions of \ourapproach maintain consistent GMACs across different bit lengths, the lower bit representations offer significant energy savings \cite{emanuel_nvidia_comp}. Figure~\ref{bubble_slurp} illustrates the energy efficiency versus \textit{F1-score} trade-offs, with bubble sizes representing model sizes in megabytes. Impressively, \ourapproach consumes $83.29\times$ less energy at the $8$-bit level compared to Whisper (base), with only a negligible $3.07\%$ drop in \textit{F1-score}. Such a dramatic reduction in energy consumption solidifies \ourapproach’s position as a highly efficient alternative to conventional SLU models.

\parlabel{FSC Dataset}
On the FSC dataset, \ourapproach achieves near-perfect \textit{F1-scores} of $99.10$–$96.12\%$ and accuracies of $99.20$–$97.39\%$, closely matching or even surpassing several SOTA baselines while maintaining a significantly smaller model size and lower computational requirements.
Compared to the most competitive SOTA models, \ourapproach requires $83$–$663\times$ less memory and $61.50\times$ fewer GMACs. Despite these drastic reductions in resource usage, \ourapproach retains exceptional accuracy, demonstrating that its compact design does not come at the expense of performance.

Figure~\ref{bubble_fsc} further illustrates the outstanding trade-off between performance and energy consumption. At the $8$-bit representation, \ourapproach consumes $3637\times$ less energy than Whisper (base) and $141\times$ less energy than Prosody + Distillation, with only a marginal \textit{F1-score} drop of $0.23\%$ and $1.14\%$, respectively. This remarkable efficiency, coupled with negligible performance degradation, underscores \ourapproach's suitability for scalable, energy-efficient SLU applications.

\subsection{Ablation Study}
\label{ablation}
We conduct an ablation study to examine the influence of model initialization (Random vs. Pre-trained) and training strategies (Distillation, Quantization after Distillation, and MCT) on the performance and efficiency of \ourapproach. The key findings are presented in Table~\ref{tab:ablation}.

\parlabel{Effect of Initialization}
Pre-trained initialization consistently outperforms random initialization across all training strategies, underscoring the critical role of leveraging prior knowledge for downstream tasks. On the SLURP dataset, distillation with pre-trained initialization achieves an \textit{F1-score} of 60.93, in stark contrast to 26.59 with random initialization—a remarkable 34.34-point improvement. This trend is even more pronounced on FSC, where all pre-trained models yield \textit{F1-scores} exceeding 96.12, demonstrating superior generalization and robustness. These results highlight that pre-training substantially accelerates convergence and enhances performance, especially for complex, real-world datasets.

\parlabel{Training Strategies}
Our results highlight that traditional distillation achieves strong performance (e.g., 98.71 F1 on FSC) but comes with significant computational overhead, resulting in large model sizes (e.g., 137.39 MB). While post-distillation quantization effectively reduces model size (e.g., SLURP: 48.13 MB vs. 78.31 MB), it severely compromises performance, leading to \textit{F1-scores} as low as 12.91.
In contrast, our MCT approach harmonizes efficiency and accuracy, delivering the best of both worlds. At 4-bit precision, \ourapproach compresses models to just 3.46 MB (SLURP) and 9.46 MB (FSC) while maintaining competitive \textit{F1-scores} of 64.39 and 96.12, respectively. This demonstrates that MCT not only mitigates the degradation typically introduced by quantization but also preserves the rich feature representations from the distillation phase, solidifying its superiority in balancing model size and performance.

\parlabel{Bit Length and Dataset Sensitivity}
Reducing bit length from 16 to 4 under MCT significantly compresses models without substantial losses in performance. For instance, on SLURP, model size drops from 13.83 MB to 3.46 MB, while \textit{F1-score} remains stable at 64.39\%. On FSC, this trend is even more pronounced: pre-trained MCT models at 16-bit precision achieve an outstanding 98.21\% F1-score, with minimal decline as bit precision decreases.
However, dataset sensitivity varies. The FSC dataset demonstrates remarkable robustness across all configurations, consistently maintaining high \textit{F1-scores} above 96.12. Conversely, SLURP exhibits greater sensitivity to extreme quantization, particularly under 4-bit constraints, suggesting that datasets with more semantic variability may require more careful tuning to maintain peak performance.

\section{Conclusion}
This study presents a unified distillation and quantization framework that achieves high performance in intent classification with minimal computational overhead. Our model attains \textit{F1-scores} of 64.39–65.07\% on SLURP and 96.12–99.10\% on FSC, with model sizes as small as 3.46 MB. Compared to state-of-the-art models, \ourapproach delivers similar accuracy with only a 2–3\% drop while significantly reducing memory and energy consumption. These results demonstrate the model’s efficiency and suitability for deployment on resource-constrained devices in real-world SLU applications.

\bibliographystyle{IEEEtran}
\bibliography{reference}

\end{document}